\documentclass{ws-procs961x669}            
\usepackage{float}
\usepackage{booktabs}
\usepackage{todonotes}
\begin{document}
\title{Estimating the Photometric Redshifts of Galaxies and QSOs Using Regression Techniques in Machine Learning}

\author{Aidin Momtaz, Mohammad Hossein Salimi and Soroush Shakeri$^*$}

\address{Department of Physics, Isfahan University of Technology,\\
Isfahan, 84156-83111, Iran\\
$^*$E-mail: s.shakeri@iut.ac.ir\\}



\begin{abstract}
Measuring distances of cosmological sources such as galaxies, stars and quasars plays an increasingly critical role in modern cosmology. Obtaining the optical spectrum and consequently calculating the redshift as a distance indicator could instantly classify these objects. As long as spectroscopic observations are not available for many galaxies and the process of measuring the redshift is time-consuming and infeasible for large samples, machine learning (ML) approaches could be applied to determine the redshifts of galaxies from different features including their photometric colors. In this paper, by using the flux magnitudes from the Sloan Digital Sky Survey (SDSS) catalog, we develop two ML regression algorithms (Decision Tree and Random Forest) for estimating the redshifts taking color indices as input features. We find that the Random Forest algorithm produces the optimum result for the redshift prediction, and it will be further improved when the dataset is limited to a subset with z $\le$ 2 giving the normalised standard deviation $\overline{\Delta Z}_{\text {norm}}=0.005$ and  the  standard deviation $\sigma_{\Delta z}=0.12$.  This work shows a great potential of using the ML approach  to determine the photometric redshifts of distant sources.


\end{abstract}


\bodymatter

\section{Introduction}\label{aba:sec1}
Spectroscopy is usually applied as a valuable technique to determine the redshift of extragalactic sources. However, its high wavelength resolution limits its accuracy, a problem that can only partially be solved with more observation time.\cite{Paul2018}
\\In recent years, a growing number of studies have relied on less precise statistical but more efficient estimates of redshifts based on broadband photometry \cite{Wilson2020,vugt2009,Mobasher2007}. A lot of key projects for the upcoming survey telescopes will involve these  photometric redshifts. The photometric redshift  is a tool used by multiple fields of astronomy to estimate the distances between objects in the sky. Since the process of redshift estimation is of great importance in various endeavors such as astronomical transient events, galaxy clustering, the mass function of the galaxy and the weak-lensing approach through constraining the presence of dark energy, several methods have been developed to choose an optimum technique\cite{Salvato2019}. These methods include leveraging training sets\cite{Collister2004}, utilizing template spectra for comparisons\cite{Benitez2000} and mostly template fitting methods\cite{Bruzual2003}. 
\\Additionally, for over a decade, cosmologists have used ML techniques based on neural networks and regression algorithms to determine photometric redshifts\cite{Zhang2019}. In an attempt to establish a quantitative approach towards the performance of Random Forests, Carliles et al. 2010 concludes that in contrast to other regression techniques, Random Forest regression overcomes several vital weaknesses\cite{Carliles2010}. Random Forest algorithm as a  non-parametric procedure does not use a statistical model to describe the underlying data. The performance of parametric methods depends on how well the model fits the underlying distribution of data\cite{Carliles2007}. However, with highly skewed noise distributions, the Random Forest still provides reliable estimate of the error distribution, and this behavior is backed up by strong theoretical support\cite{Baldeschi2021}.
\\In Hoyle et al. (2015), the ML architecture Decision Trees were applied to the photometric redshift estimation as an analysis of feature importance selection. Such an investigation in this regard remarked  in five optical frequency bands known as u-g-r-i-z, the flux magnitudes available from the SDSS, as a determinative input feature that encodes most of the information about the redshifts of galaxies or quasars\cite{Hoyle2015}.
\\Over the last 20 years, the SDSS has made a map of the universe. SDSS measurements of the galaxies, quasars and intergalactic gas structure have contributed significantly to tests of the standard cosmological model that describes our understanding of the history and future of the universe. Data Release 16 of SDSS \cite{Ahumada2020}includes infrared, extragalactic and integral field spectra for nearby galaxies. The survey has mapped 4846156 useful spectra, 2863635 of which are galaxies and 960678 are quasi-stellar objects or the so-called quasars.

In this paper,  we  focus on  Data Release 16 of SDSS and  design  a  Decision  Tree  and  a  Random  Forest  algorithm  to obtain  an  accurate  estimation  of  the  redshifts  and  then  investigated  their evaluation procedures to obtain optimum algorithms. We create a dataset of color indices acting as an approximation for the spectrum and  as  our  input  features. A noteworthy aspect of this work is the effort put into incorporating more information than a simple point estimate of the redshift. We determine the uncertainty associated with redshift estimates and  calculate a posterior distribution\cite{Oyaizu2008}.

The rest of the paper is organized as follows. In Sec. \ref{sec2}, we explain ML algorithms including Decision Trees and Random Forests. In Sec. \ref{sec3}, we introduce the 16th data release of SDSS (DR16) 
and our input features besides our ML methodology for training and learning. The analysis and results are presented in Sec. \ref{analysis} where we apply optimizing processes for Decision Tree and Random Forest algorithms and compare the results. Finally, we discuss our results in Sec. \ref{conclusion}.

\section{Machine Learning Algorithms}
\label{sec2}

\subsection{Decision Trees}
A Decision Tree is an ML algorithm which is used for both classification and regression learning tasks \cite{Breiman:1984jka}. Based on a set of input features, a Decision Tree generates its corresponding output targets. This is accomplished by a series of single decisions, each representing a node or branching of the tree. Following the training data, the Decision Tree learning algorithm determines which decision should be made at each branch. Various metrics are employed by each algorithm in order to determine what is the best way to split the data (e.g. Gini impurity or information gain). 
\begin{figure}[h]
\begin{center}
\includegraphics[width=5in]{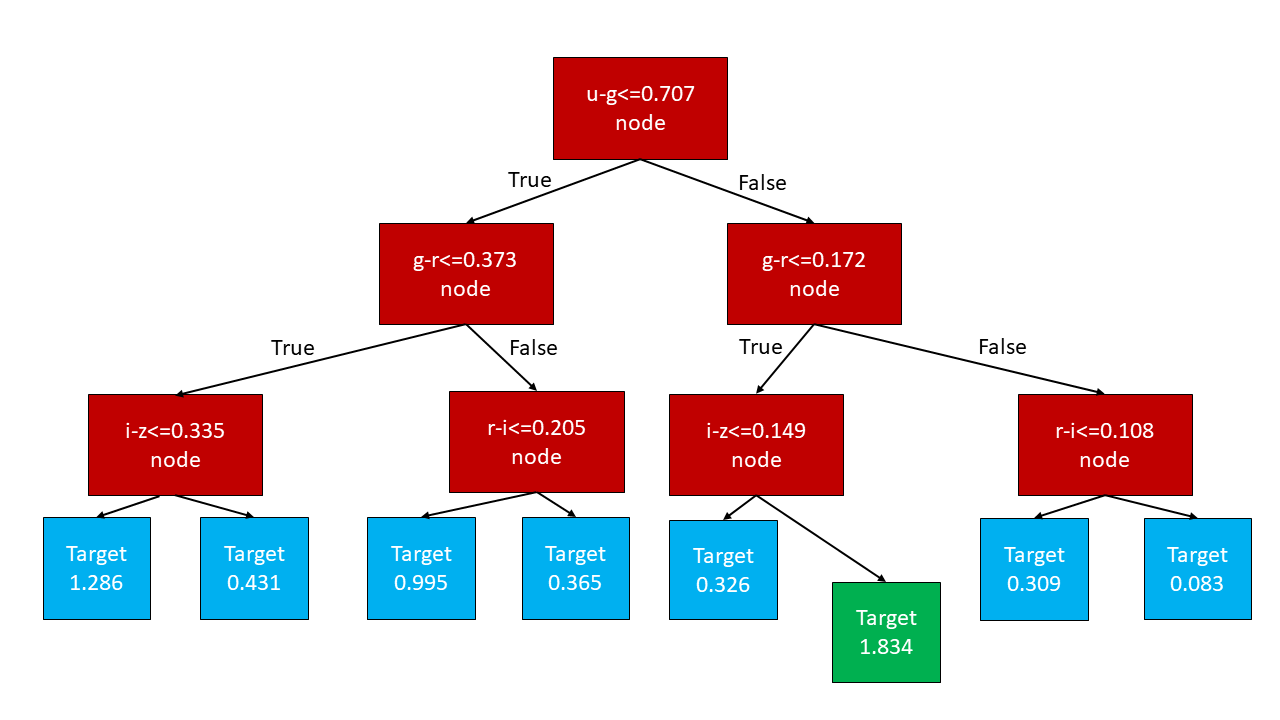}
\end{center}
\caption{Schematic of the Decision Tree algorithm.}
\label{aba:fig1}
\end{figure}
\\The inputs of the calculated color indices into a series of decision nodes are shown above (Fig. \ref{aba:fig1}), and  the target redshift is calculated as an output through the series of decision nodes.
To implement ML, we will use Python's scikit-learn library. Using scikit-learn Decision Tree regression, a set of input features and target values are taken into account and then the model is constructed to adapt to new data.
\subsubsection{Important Hyperparameters of Decision Trees}
The most significant hyperparameter of Decision Trees is the \textit{max-depth}. In all other cases, the nodes are expanded until all leaves are pure. In the case of a continuous value, as it is in this study, the most commonly used criteria to determine split locations are Mean Square Error, Poisson deviance as well as Mean Absolute Error.
\subsection{Random Forests}\label{sec:sec2.2}

\begin{figure}[h]
\begin{center}
\includegraphics[width=5in]{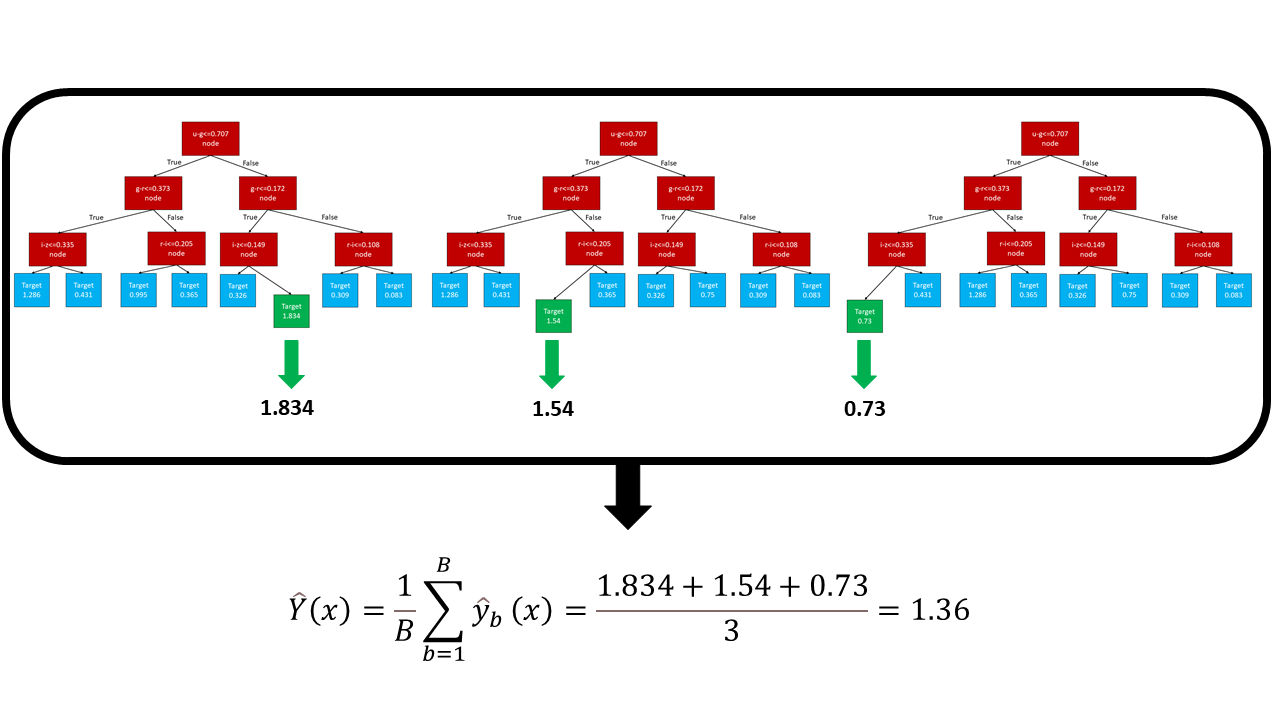}
\end{center}
\caption{Schematic of the Random Forest algorithm.}
\label{aba:rf}
\end{figure}

Random Forests can be viewed as an ensemble of Decision Trees in which random subsamples of the input attributes are used to create each tree (Fig. \ref{aba:rf}) \cite{Breiman:2001hzm}. In this case, the trees will only learn a portion of the input attribute pattern, hence they will be poor classifiers. \\Additionally, the trees learn only a part of the data, so they cannot learn artificial structures or be influenced by correlated attributes the same way as a neural network can. As a final output, they give an average result based on the trained trees, which indicates the probability of each object belonging to one of the specified categories. This method gathers attractive advantages. For instance, the model resists overfitting and is robust against correlated input attributes. In most cases, Random Forest produces excellent results. Moreover, it is one of the most common algorithms, owing to its simplicity and versatility.
\subsubsection{Important Hyperparameters of Random Forests}
First and foremost, there is the \textit{n-estimators} hyperparameter, which is essentially the number of trees the algorithm builds before taking the maximum voting or calculating the average prediction. It is generally true that a higher number of trees leads to a better estimation and helps to be more accurate, but it can also slow down the calculation speed.
\section{Data}
\label{sec3}
In this study, SDSS Data Release 16 \cite{Ahumada2020} is used as the data source. Observations for the SDSS have been carried out from Apache Point Observatory (APO) since 1998 (using the 2.5m Sloan Foundation Telescope\cite{Gunn2006}) and from Las Campanas Observatory (LCO) since 2017 (using the du Pont 2.5m Telescope).
\\The newest DR16 contains the final sets of spectra collected as part of the main eBOSS observing program. SDSS DR16, then, ends a twenty-year stretch of performing a large-scale survey of the structure of the universe. SDSS has produced the largest catalog of spectroscopic redshifts of galaxies compare to any other program over this time period. It is worth mentioning that DR16 provides spectra with usable redshifts for around 2.6 million unique galaxies. The photometric properties of around 100 million galaxies were also measured in this dataset.
\\DR16 includes SDSS data products that are freely available through several channels. 
A large number of photometrically selected galaxies with spectroscopic redshifts are present in the SDSS that lend themselves well to the analysis presented in this paper to use as training, cross-validation and test samples.
\subsection{Input Features}
\begin{figure}[h]
\begin{center}
\includegraphics[width=4in]{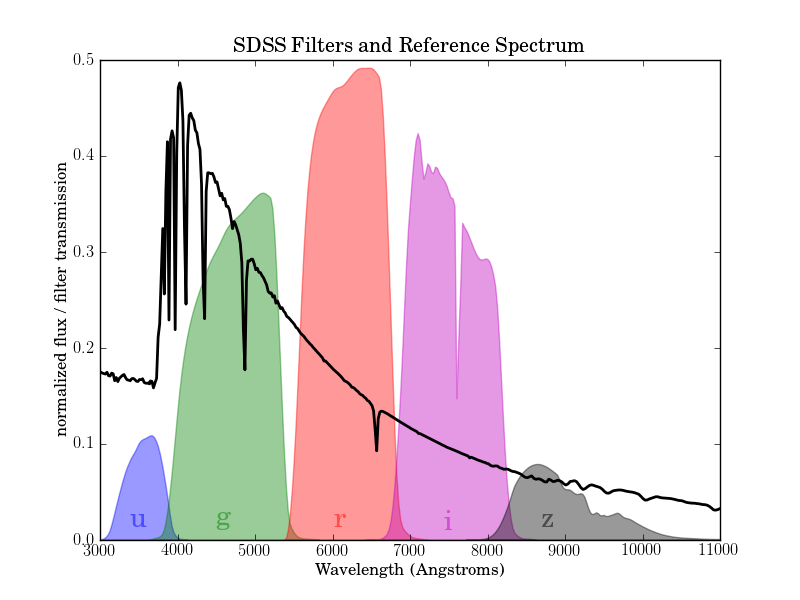}
\end{center}
\caption{SDSS filters and reference spectrum \cite{astroMLText}.}
\label{uuuuu}
\end{figure}

Using the color indices as our input features and the photometric redshift as the output, we have constructed a Decision Tree and  a Random Forest. The data that we use for training are collected through accurate spectroscopic measurements of SDSS. These color indices have been created from flux magnitudes which are the total flux received in five frequency bands known as u-g-r-i-z (Fig. \ref{uuuuu}). Also, an astronomical color is derived from the difference in magnitudes of two filters, i.e. u - g. An object's color index provides an approximation to its spectrum, allowing it to be classified into various types.
\\ We can get information about the physical processes taking place in distant astronomical sources by precise  analysis of high-resolution spectra. Since the process of obtaining these spectra is very time-consuming and costly, astrophysicists usually observe objects through broadband filters and record them using the magnitude system. As an example, we can define the magnitude in the u-band filter above as follows:
\begin{equation}
        u=m_{\text {ref }}-2.5 \log {10}\left[\int_{0}^{\infty} F(\lambda) S(\lambda) d \lambda\right],
    \end{equation}
where $F(\lambda)$ is the star’s flux  at wavelength $\lambda$, $S(\lambda)$ is a sensitivity function describes the fraction of the star’s flux that is detected at a specific wavelength and $m_{\text {ref }}$ is apparent bolometric magnitude.

\subsection{Training and Learning}
Following the conventional ML methodology, the galaxy catalog is then subdivided into training and testing samples, with portions of 80 and 20,  respectively. For each architecture and hyperparameter set, the ML system is trained using a training sample. Testing the learned machine's generalization ability based on a test sample is necessary to determine whether it truly generalizes to new datasets. Furthermore, k-fold cross-validation allows us to test the accuracy of our model. Our model is trained k times, with each training test recording the accuracy. We train the model every time using a different combination of k-1 subsets, and we test it with the final kth subset. Then, the overall accuracy of the model is calculated by taking the average of the k accuracy measurements.
\section{Analysis and Results}
\label{analysis}
\begin{figure}[H]
\begin{center}
\includegraphics[width=4in]{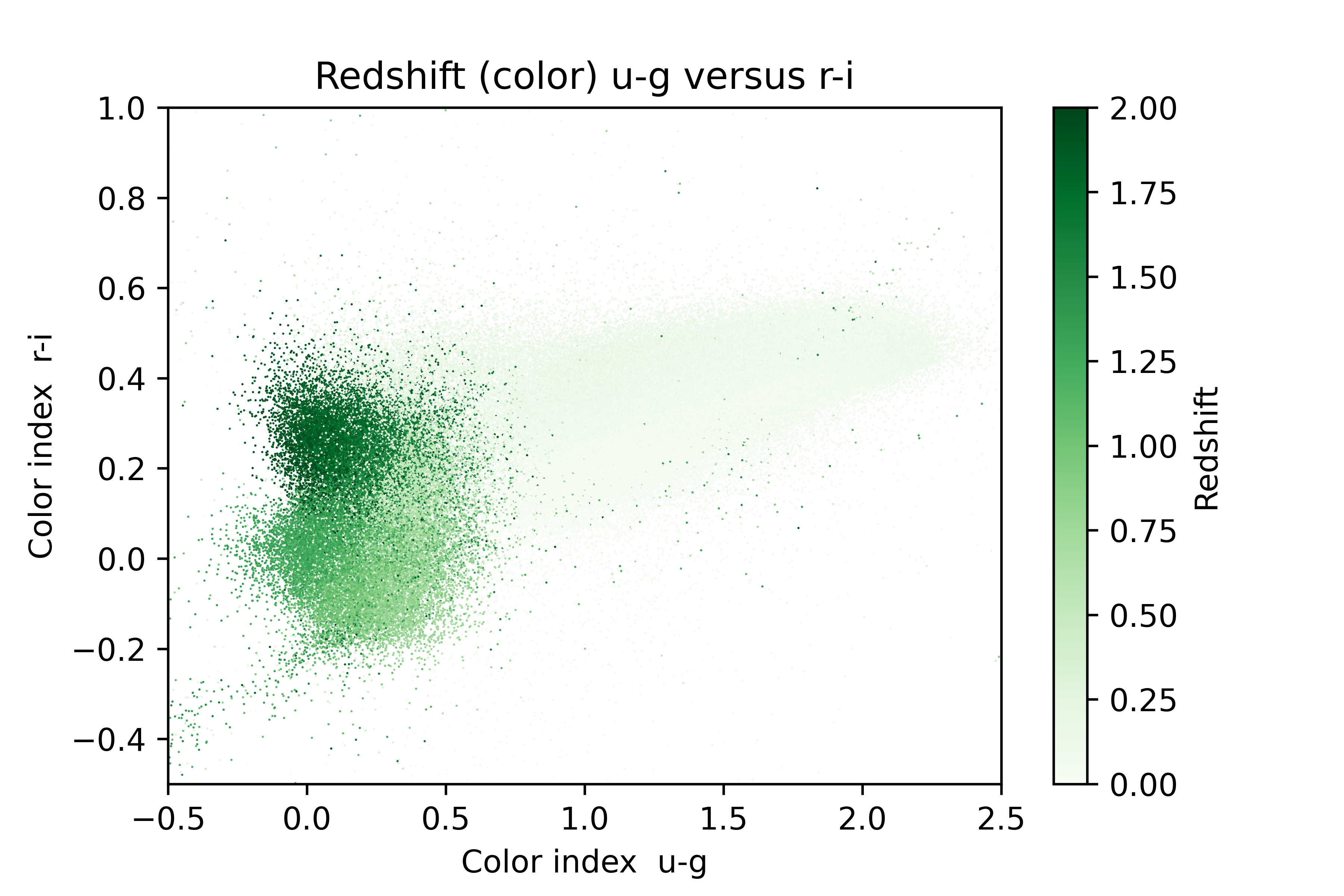}
\end{center}
\caption{Contour map of the redshifts based on a combination of color indices.}
\label{countour}
\end{figure}

In the following, we start to analyze the final results of our structured algorithms. First we make a contour map of the redshifts based on a combination of color indices (Fig. \ref{countour}), the results indicate that relatively well-defined regions of similar redshifts can be extracted and therefore the redshift of a new data point can be inferred from the color indices. So, this gesture underlines the fact that color indices are, indeed, appropriate input features for our ML algorithms.
In Fig. \ref{fig5}, the redshift distribution of galaxies and quasars are presented. In our dataset  there are more galaxies than quasars, and galaxies usually have lower redshifts. 
\def\figsubcap#1{\par\noindent\centering\footnotesize(#1)}
\begin{figure}[h]%
\begin{center}
  \parbox{2.1in}{\includegraphics[width=2.5in]{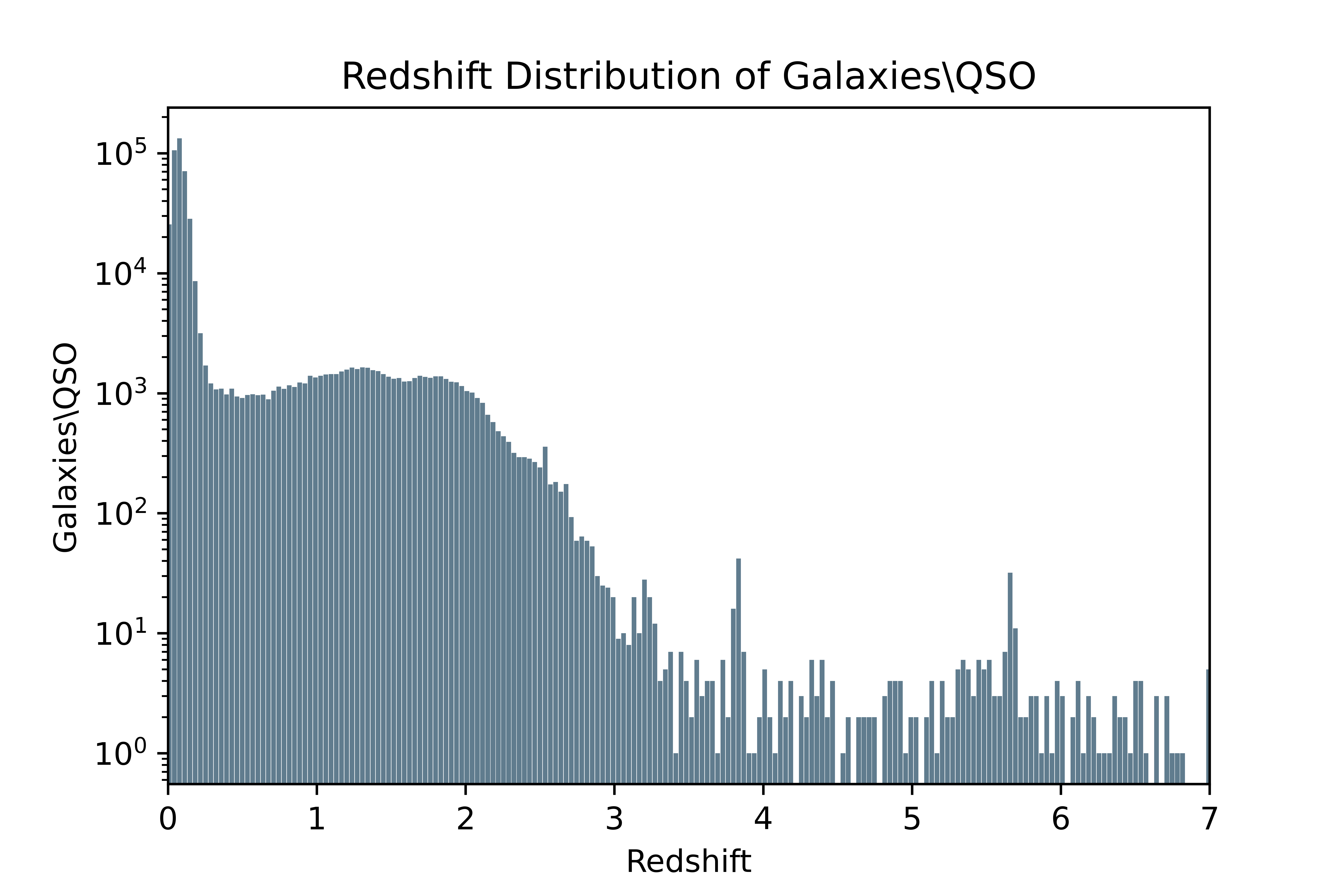}\figsubcap{a}}
  \hspace*{4pt}
  \parbox{2.1in}{\includegraphics[width=2.5in]{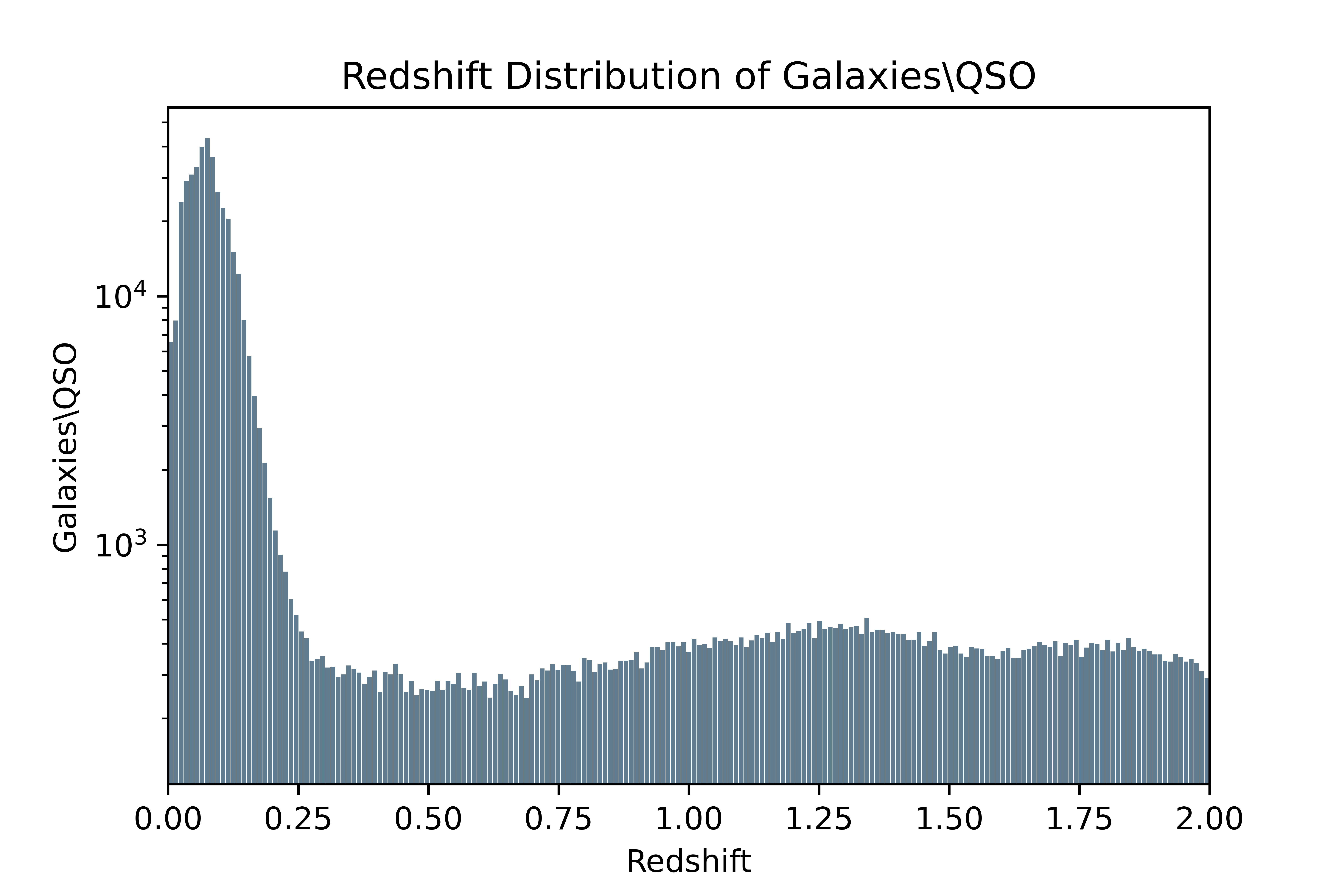}\figsubcap{b}}
  \caption{Redshift distribution of Galaxies/QSOs. (a) Total range of the dataset. (b) limited range (z$\le$2) of the SDSS dataset.}%
  \label{fig5}
\end{center}
\end{figure}
\\We divide our dataset  into two ranges, i) contains all the measured redshifts ii) includes those below value 2. The reason behind  this procedure is to achieve better performance of the algorithms, which will be discussed in detail in the following sections.
\subsection{Optimizing Process for Decision Trees}
There are some limitations for Decision Tree algorithms, including an over-fitting tendency. This means that it would potentially create a tree that is too complicated and does not address the statistical outliers in the data. In the process, general trends may not be accurately characterized.
\\Among the reasons for the over-fitting is that the algorithm works by trying to optimize each node's decision locally. During our analysis, we will examine the impact of constraining the number of decision node rows (tree depth) on predictions. Various tree depths can be used to investigate whether the tree is over-fitting or not. We are particularly interested in comparing the algorithm performance on test data to its performance on training data (Fig. \ref{analysisofthemax}) 
\def\figsubcap#1{\par\noindent\centering\footnotesize(#1)}
\begin{figure}[H]%
\begin{center}
  \parbox{2.1in}{\includegraphics[width=2.5in]{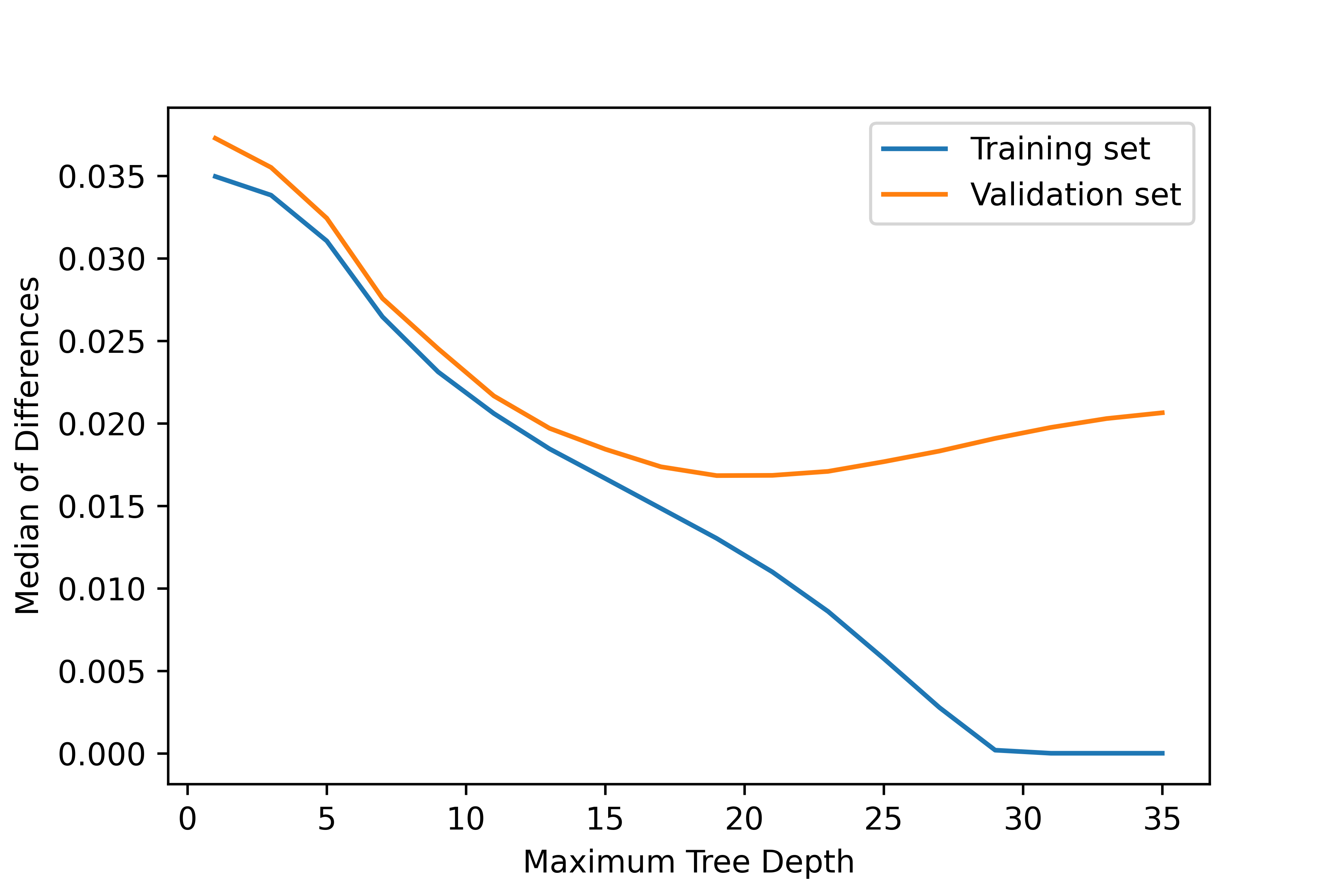}\figsubcap{a}}
  \hspace*{4pt}
  \parbox{2.1in}{\includegraphics[width=2.5in]{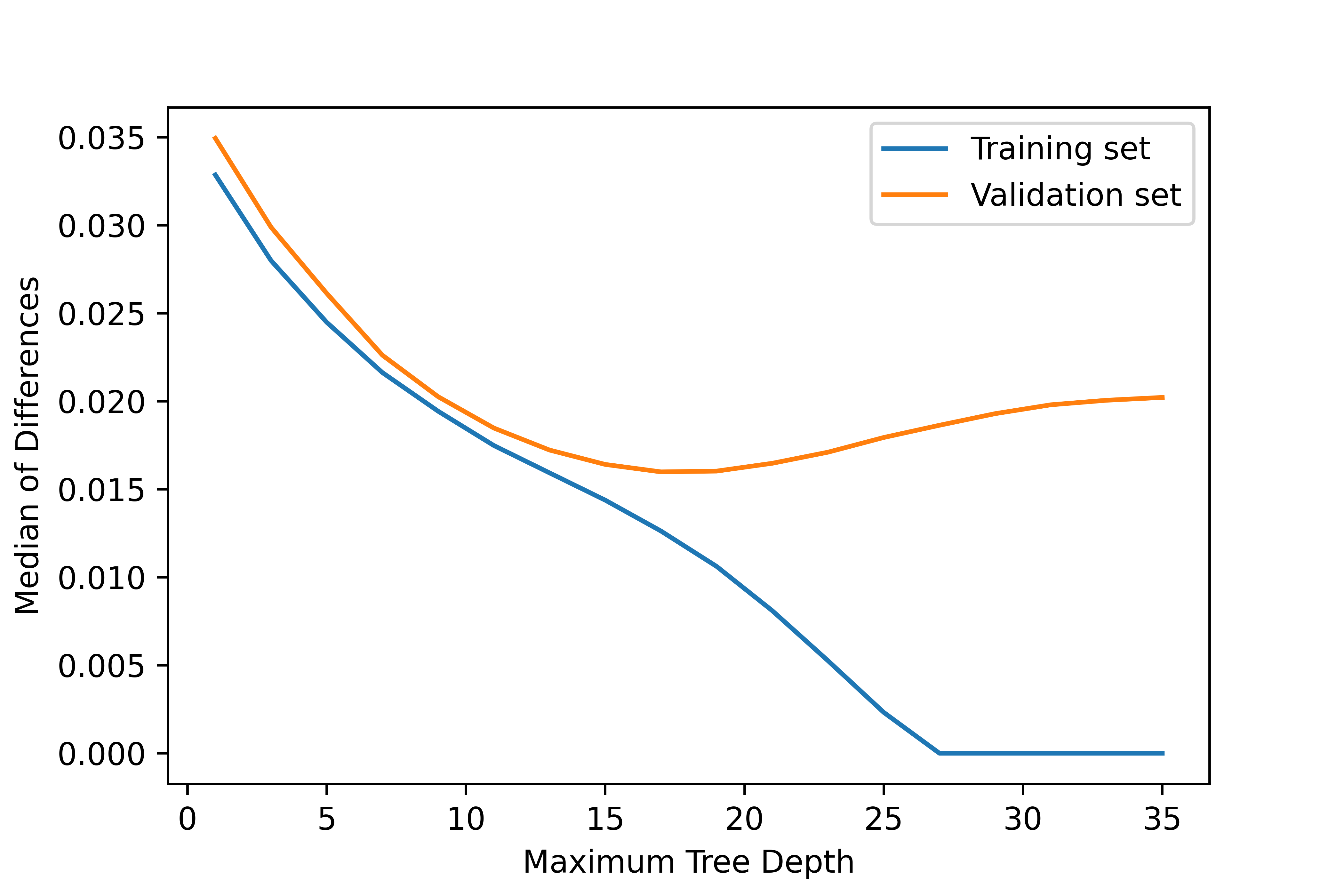}\figsubcap{b}}
  \caption{Analysis of the maximum tree depth (a) Total range of the dataset. (b) limited range (z$\le$2) of the SDSS dataset.}%
  \label{analysisofthemax}
\end{center}
\end{figure}

\subsection{Optimizing Process for Random Forests}
Creating a parameter grid prior to fitting is the first step in optimizing the Random Forest algorithm. A new combination of features is selected on each iteration, especially the \textit{n-estimators} parameter, which represents the number of trees. In our Random Forest algorithm, we see that 600 trees will give the best result, as measured by the median value of the residual of measured and predicted redshifts. By setting the \textit{n-estimators} to 600, we investigate the optimum amount of each tree's depth as in our prior grid search in the algorithm.

\subsection{Illustrations of the Models Performance}\label{sec4.3.}
\begin{figure}[H]
\begin{center}
\includegraphics[width=5in]{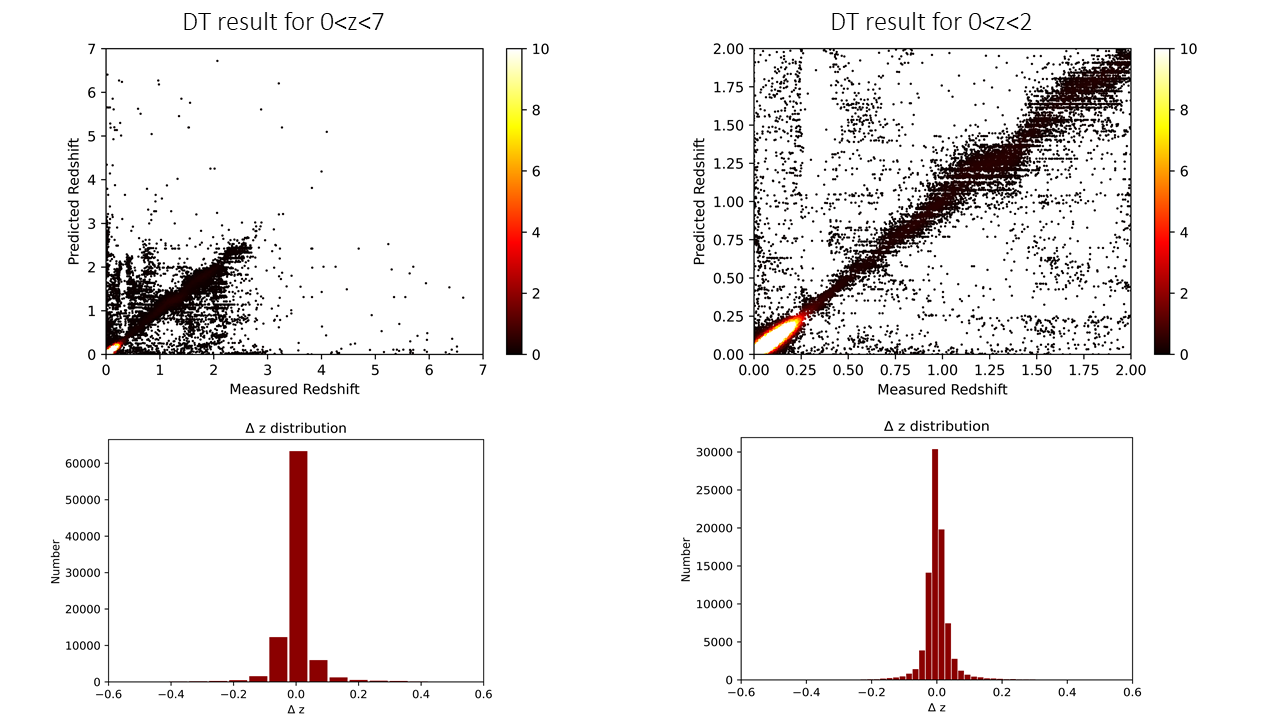}
\end{center}
\caption{The predictions of the Decision Tree algorithm for two ranges of redshifts  in SDSS dataset.}
\label{dtdeltaz}
\end{figure}

\begin{figure}[H]
\begin{center}
\includegraphics[width=5in]{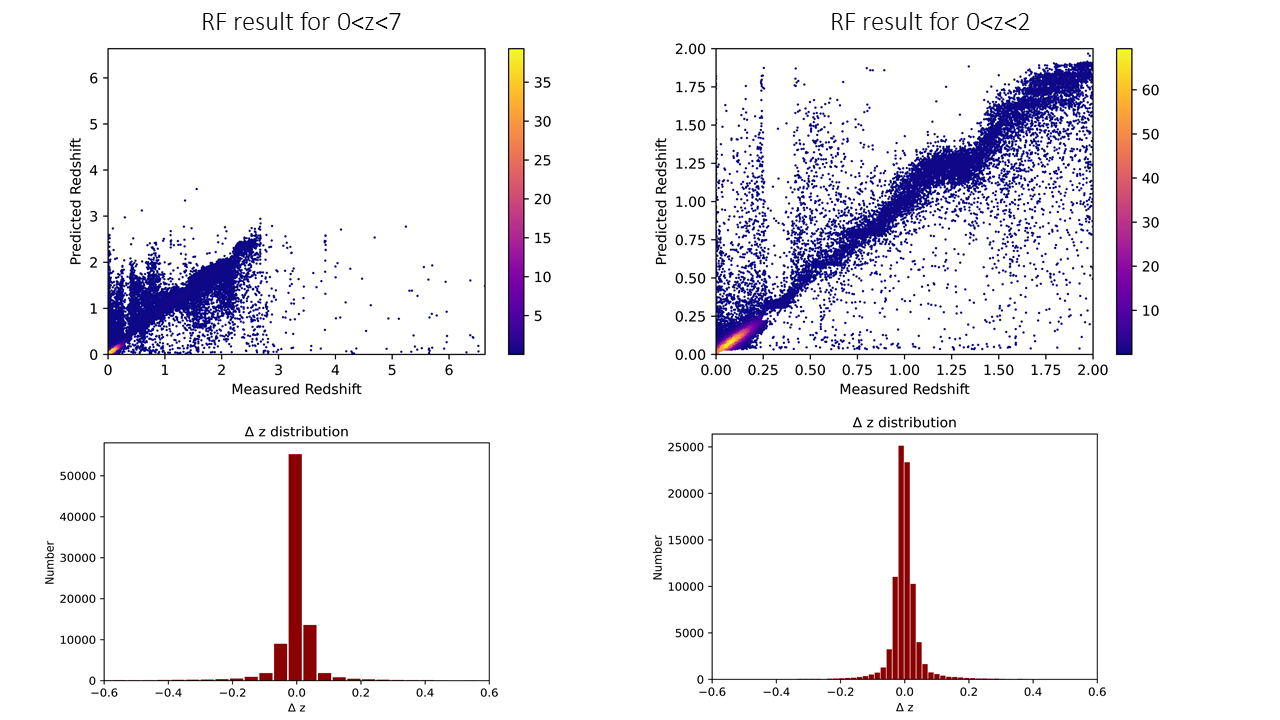}
\end{center}
\caption{The predictions of the Random Forest algorithm for two ranges in SDSS dataset.}
\label{rfdeltaz}
\end{figure}

Once the maximum tree depths have been determined, the algorithms are ready to take on the two different sets of data extracted from DR16. We have used 450000 of the catalog's data based on our system's performance. There are 400000 data under the value 2 in the parts of the paper where calculations has been executed on the filtered dataset. Figs. \ref{dtdeltaz} and  \ref{rfdeltaz} present the measured redshifts from the survey versus the algorithms predicted redshifts. The colorbars indicate the density of the galaxies or quasars in the datasets. A straight line with an angle of 45 degrees  illustrates the success of the algorithms where measured redshifts is equal to the predicted ones.
The overall performance of each algorithm was evaluated using three metrics \cite{2021}: model accuracy, the Root Mean Square Error (RMSE), and the normalized standard deviation. The normalized standard deviation is defined by:
\begin{equation}
\Delta z(\text {norm}) \equiv \frac{z_{\mathrm{spec}}-z_{\mathrm{phot}}}{z_{\mathrm{spec}}+1}.
\label{eq2}
\end{equation}
We calculate the standard deviation of the photometric redshifts from the spectroscopic redshift or namely the RMSE:
\begin{equation}
\sigma_{\Delta z}=\sqrt{\frac{1}{N} \sum_{i=1}^{N} \Delta z^{2}}.
\end{equation}
Fig. \ref{dtdeltaz} illustrates how the Decision Tree algorithm has performed on both subsets of the dataset. Each datum's predicted redshift is plotted as a function of its spectroscopically measured value. The left panel shows the the results for the whole range of the redshift data, while the right panel is limited to the results with redshifts below the value 2. Moreover, the error distributions of the predicted redshifts appear under each graph. We provide the mathematical criteria to identify the performance of each model in  Tables \ref{rfdeltaz} and \ref{tbl2}. As it is seen in both of the diagrams,  limiting the redshift to values below 2 leads to more accurate results.

As the Table \ref{tbl1} shows, filtering data into lower redshifts leads to a significant improvement in the model's performance. The accuracy of the algorithm has increased by 15\%, the median difference, the normalized error as well as the standard deviation were all decreased by corresponding amounts. The explanation which underlies this improvement could be the fact that by splitting the dataset, there are no more considerable biases in the values of target redshifts and the algorithm would not experience a scattered distribution of values in the training procedure. Thereby, there should be no excessive miscalculation or poor estimation in the testing mode.
\begin{table}[!htbp]
\tbl{Analytical overview of statistical parameters for the Decision Tree algorithm for two ranges of redshifts z$\le$7 and z$\le$2. }
{\begin{tabular}{cccc}
\toprule
Parameters & Decision Tree & Decision Tree \\
& (z$\le$7) & (z$\le$2) \\
\colrule
Accuracy\hphantom{00} & \hphantom{0}70.17\% & \hphantom{0}85.26\%  \\
Max Depth\hphantom{00} & \hphantom{0} 19 & \hphantom{0} 17\\
Median Difference & 0.017 & 0.0156\hphantom{0}\\
$\overline{\Delta Z}_{\text {norm}}$ & 0.0135 & 0.005\hphantom{0}\\
$\sigma_{\Delta z}$ & 0.28 & 0.16\hphantom{0}\\
\botrule
\end{tabular}
}
\label{tbl1}
\end{table}

Contemplating what was discussed in Sec.\ref{sec:sec2.2}, as the number of trees grow, the accuracy of the predicted values increases. Thus, it is expected  that the Random Forests would have a better overall performance in terms of making more accurate predictions. As shown in Fig. \ref{rfdeltaz}, both in the scattered plot and the distribution plot, the amount of precise estimations has increased. Since the colorbar range indicates the density of data in the diagram and it includes a greater range of numbers than the Decision Tree algorithm. We  conclude that the Random Forest algorithm has better accuracy. This fact is also presented in terms of statistical parameters such as the standard deviation as shown in Table \ref{tbl2}.

\begin{table}[!htbp]
\tbl{Analytical overview of statistical parameters for the Random Forest algorithm for two ranges of redshifts z$\le$7 and z$\le$2.}
{\begin{tabular}{@{}cccc@{}}
\toprule
Parameters & Random Forest & Random Forest \\
& (z$\le$7) & (z$\le$2) \\
\colrule
Accuracy\hphantom{00} & \hphantom{0}81.02\% & \hphantom{0}91.00\%  \\
Max Depth\hphantom{00} & \hphantom{0}25 & \hphantom{0}13\\
Number of Trees\hphantom{0} & 600 & 600\hphantom{0} \\
Median Difference & 0.0164 & 0.0154\hphantom{0}\\
$\overline{\Delta Z}_{\text {norm}}$ & 0.013 & 0.005\hphantom{0}\\
$\sigma_{\Delta z}$ & 0.23 & 0.12\hphantom{0}\\
\botrule
\end{tabular}
}
\label{tbl2}
\end{table}
 
\section{Conclusion and Remarks}
\label{conclusion}
The current research addresses two leading ML regression algorithms namely the Decision Tree and the Random Forest, that were structured for estimating the redshifts of distant objects such as galaxies and quasars. Since there exist vital limitations in obtaining spectroscopic measurements, the photometric optical band data are widely investigated in this realm. This paper uses the 16 Data Release of the SDSS, where photometric colors are used as input features of the mentioned models.
After developing the Decision Tree and the Random Forest algorithms and evaluating the final results, it was concluded that the Random Forest algorithm will significantly perform better in this case. The Random Forest algorithm leads to a much better accuracy and marginally better standard deviation and median difference as seen in Table \ref{tbl3}. 
\\Another noteworthy improvement was obtained by filtering the dataset to redshifts below the value 2. The significant effect of redshift splitting  was illustrated in Sec. \ref{sec4.3.} and we discussed about the possible reasons behind it.
\begin{table}[!htbp]
\tbl{Comparison between statistical parameters of the Decision Tree and the Random Forest algorithms for  z$\le$2.}
{\begin{tabular}{@{}cccc@{}}
\toprule
Parameters & Decision Tree & Random Forest \\
& (z$\le$2) & (z$\le$2) \\
\colrule
Accuracy\hphantom{00} & \hphantom{0}85.26\% & \hphantom{0}91.00\%  \\
Median Difference & 0.0156 & 0.0154\hphantom{0}\\
$\overline{\Delta Z}_{\text {norm}}$ & 0.005 & 0.005\hphantom{0}\\
$\sigma_{\Delta z}$ & 0.16 & 0.12\hphantom{0}\\
\botrule
\end{tabular}
}
\label{tbl3}
\end{table}

A few recent attempts for redshift estimation 
applying similar developed ML algorithms are addressed in Table \ref{tbl4}. The crucial factor affecting their evaluations is the amount of data in the training sets. Remarkably, in this work taking 400000 data into consideration for the training set, we achieved nearly equivalent accuracy compared to previous works. 
\begin{table}[!htbp]
	\tbl{Comparison among results obtained  using similar methodologies by different collaborations and our results in this work. }
	{\begin{tabular}{ccccc}
			\toprule
			Refrence Articles   & Data Release & Training Set   & $\overline{\Delta Z}_{\text {norm}}$   & ML Algorithm  \\
			\midrule
			Beck(2016)       & 12    & 1,976,978    &  $5.84\times 10^{-5}$   & Local Linear Regression  \\
			Paul(2018) \cite{Paul2018}  & 12        & 20,000     & $2\times 10^{-3}$   & Random-Forest  \\
			Baldeschi(2021) \cite{Baldeschi2021}        & 16     & 1,251,249    & $1\times 10^{-3}$   & Random-Forest  \\
			This work   & 16     & 320,000     & $5\times 10^{-3}$   & Random-Forest \\
			\bottomrule
		\end{tabular}
	}
	\label{tbl4}
\end{table}


Our results show the great potential of the ML methods for redshift  estimation of distant sources using color index features. It is worth mentioning that applying ML methods is unavoidable when a large number of astrophysical data will be obtained from the next generation of sky surveys as the era of big data has started.

\section*{Acknowledgments}
The article is prepared for the proceedings of the sixteenth Marcel Grossmann meeting (MG16). The authors would like to thank the organizers and conveners of the MG16 for giving the opportunity to present this work and useful discussions during the meeting.

\bibliographystyle{ws-procs961x669}
\bibliography{ws-pro-sample}

\begin{thebibliography}{10}

\bibitem{Paul2018}
N.~Paul, N.~Virag and L.~Shamir, {A catalog of photometric redshift and the
  distribution of broad galaxy morphologies}, {\em Galaxies} {\bf 6}, 1
  (2018).

\bibitem{Wilson2020}
D.~Wilson, H.~Nayyeri, A.~Cooray and B.~H{\"{a}}u{\ss}ler, {Photometric
  Redshift Estimation with Galaxy Morphology Using Self-organizing Maps}, {\em
  The Astrophysical Journal} {\bf 888}, p.~83  (2020).

\bibitem{vugt2009}
J.~v. Vugt, {Photometric Redshift Estimation of Distant Quasars}  (2016).

\bibitem{Mobasher2007}
B.~Mobasher, P.~Capak, N.~Z. Scoville, T.~Dahlen, M.~Salvato, H.~Aussel, D.~J.
  Thompson, R.~Feldmann, L.~Tasca, O.~Lefevre, S.~Lilly, C.~M. Carollo, J.~S.
  Kartaltepe, H.~McCracken, J.~Mould, A.~Renzini, D.~B. Sanders, P.~L.
  Shopbell, Y.~Taniguchi, M.~Ajiki, Y.~Shioya, T.~Contini, M.~Giavalisco,
  O.~Ilbert, A.~Iovino, V.~{Le Brun}, V.~Mainieri, M.~Mignoli and M.~Scodeggio,
  {Photometric Redshifts of Galaxies in COSMOS}, {\em The Astrophysical Journal
  Supplement Series} {\bf 172}, 117  (2007).

\bibitem{Salvato2019}
M.~Salvato, O.~Ilbert and B.~Hoyle, {The many flavours of photometric
  redshifts}, {\em Nature Astronomy} {\bf 3}, 212  (2019).

\bibitem{Collister2004}
A.~Collister and O.~Lahav, { ANN z : Estimating Photometric Redshifts Using
  Artificial Neural Networks }, {\em Publications of the Astronomical Society
  of the Pacific} {\bf 116}, 345  (2004).

\bibitem{Benitez2000}
N.~Benitez, {Bayesian Photometric Redshift Estimation}, {\em The Astrophysical
  Journal} {\bf 536}, 571  (2000).

\bibitem{Bruzual2003}
G.~Bruzual and S.~Charlot, {Stellar population synthesis at the resolution of
  2003}, {\em Monthly Notices of the Royal Astronomical Society} {\bf 344},
  1000  (2003).

\bibitem{Zhang2019}
K.~Zhang, D.~J. Schlegel, B.~H. Andrews, J.~Comparat, C.~Sch{\"{a}}fer, J.~A.
  {Vazquez Mata}, J.-P. Kneib and R.~Yan, {Machine-learning Classifiers for
  Intermediate Redshift Emission-line Galaxies}, {\em The Astrophysical
  Journal} {\bf 883}, p.~63  (2019).

\bibitem{Carliles2010}
S.~Carliles, T.~Budav{\'{a}}ri, S.~Heinis, C.~Priebe and A.~S. Szalay, {Random
  forests for photometric redshifts}, {\em Astrophysical Journal} {\bf 712},
  511  (2010).

\bibitem{Carliles2007}
S.~Carliles, T.~Budav{\'{a}}ri, S.~Heinis, C.~Priebe and A.~Szalay,
  {Photometric Redshift Estimation on SDSS Data Using Random Forests},  {\bf
  XXX}, 1  (2007).

\bibitem{Baldeschi2021}
A.~Baldeschi, M.~Stroh, R.~Margutti, T.~Laskar and A.~Miller, {Photometric
  redshift estimation of galaxies in the P\lowercase{an}-STARRS 3$\pi$ survey-
  I. Methodology}  (2021).

\bibitem{Hoyle2015}
B.~Hoyle, M.~{Michael Rau}, R.~Zitlau, S.~Seitz and J.~Weller, {Feature
  importance for machine learning redshifts applied to SDSS galaxies}, {\em
  Monthly Notices of the Royal Astronomical Society} {\bf 449}, 1275  (2015).

\bibitem{Ahumada2020}
R.~Ahumada {\em et~al.}, {The 16th Data Release of the Sloan Digital Sky
  Surveys: First Release from the APOGEE-2 Southern Survey and Full Release of
  eBOSS Spectra}, {\em Astrophys. J. Suppl.} {\bf 249}, p.~3  (2020).

\bibitem{Oyaizu2008}
H.~Oyaizu, M.~Lima, C.~E. Cunha, H.~Lin and J.~Frieman, {Photometric Redshift
  Error Estimators}, {\em The Astrophysical Journal} {\bf 689}, 709  (2008).

\bibitem{Breiman:1984jka}
L.~Breiman, J.~Friedman, R.~A. Olshen and C.~J. Stone, {\em {Classification and
  regression trees}} (Chapman and Hall/CRC, 1984).

\bibitem{Breiman:2001hzm}
L.~Breiman, {Random Forests}, {\em Machine Learning} {\bf 45}, 5  (2001).

\bibitem{Gunn2006}
J.~E. Gunn {\em et~al.}, {The 2.5 m Telescope of the Sloan Digital Sky Survey},
  {\em Astron. J.} {\bf 131}, 2332  (2006).

\bibitem{astroMLText}
{\v Z}.~{Ivezi{\'c}}, A.~{Connolly}, J.~{Vanderplas} and A.~{Gray}, {\em
  Statistics, Data Mining and Machine Learning in Astronomy} (Princeton
  University Press, 2014).

\bibitem{2021}
S.~J. Curran, J.~P. Moss and Y.~C. Perrott, Qso photometric redshifts using
  machine learning and neural networks, {\em Monthly Notices of the Royal
  Astronomical Society} {\bf 503}, p. 2639–2650 (Feb 2021).

\end{thebibliography}

\end{document}